\documentclass[aps,prb,amsmath,amssymb,footinbib,showpacs,twocolumn,superscriptaddress,longbibliography]{revtex4-2}
\usepackage{amsmath}
\usepackage{amssymb}
\usepackage{amsthm}
\usepackage{balance}
\usepackage{setspace}
\usepackage{graphicx}
\usepackage{dsfont}
\allowdisplaybreaks[4]
\usepackage{braket}
\usepackage{mathrsfs}
\usepackage{comment}
\usepackage{float}
\usepackage[colorlinks = true,linkcolor = red,urlcolor  = blue,citecolor = blue,anchorcolor = blue]{hyperref}
\usepackage[utf8]{inputenc}
\usepackage[english]{babel}
\usepackage{commath}
\usepackage{xcolor}

\usepackage{bm}
\usepackage[caption=false]{subfig}

\begin{document}
\title{Flat-band systems: A platform to realize the moments of Quantum Metric}

\author{Rishi Gangadhar}
\affiliation{Department of Physics and Astronomy, Clemson University, Clemson, South Carolina 29634, USA}

\author{Binayyak Roy}
\affiliation{Department of Physics and Astronomy, Clemson University, Clemson, South Carolina 29634, USA}

\author{Vito Scarola}
\affiliation{Department of Physics, Virginia Tech, Blacksburg, Virginia 24061, USA}

\author{Sumanta Tewari}
\affiliation{Department of Physics and Astronomy, Clemson University, Clemson, South Carolina 29634, USA}

\begin{abstract}

The quantum metric is a fundamental geometric property of Bloch bands that governs a wide range of physical phenomena, yet experimentally accessing it remains challenging. We show that the momentum-space variation of the quantum metric can be probed through the real-space dynamics of a Gaussian wave packet prepared in an isolated flat band. Following the application of a weak constant force, we demonstrate analytically that the leading order change in the wave-packet variance is linear in time and contains a contribution proportional to the wave-packet averaged gradient of the quantum-metric trace. This geometric contribution remains finite despite the absence of band dispersion. We illustrate the effect using a gapped Lieb lattice with an exactly flat middle band and vanishing Berry curvature, allowing the quantum-metric contribution to the spreading to be isolated. The numerical results agree with the analytical prediction and confirm the quantum-metric contribution to the wave-packet spreading. The required lattice preparation, tunable hopping amplitudes, applied force, and density measurements are compatible with established ultracold-atom techniques, providing a possible platform for observing the predicted response. Our results establish wave-packet spreading as a dynamical probe of quantum geometry in flat-band systems.

\end{abstract}

\maketitle

\section{Introduction}

Flat-band systems are an emerging class of physical systems characterized by one or more nearly dispersionless bands, for which the band energy $E_n(\bm{k})$ is independent of, or only weakly dependent on, the crystal momentum $\bm{k}$~\cite{Leykam2018,Rhim2019FlatBands}. Since a flat band has little or no dispersion, the usual kinetic-energy contribution is strongly suppressed. As a result, interactions, disorder, external fields, and coupling to nearby bands can strongly influence the low-energy physics even when their energy scales are smaller than the underlying hopping amplitude~\cite{Mielke1991,Tasaki1992,Huber2010FlatBands,Lin2014FlatbandOptical,Leykam2018}. Consequently, flat bands provide a promising platform for studying disorder-induced transport and responses governed by the quantum geometry of the Bloch states~\cite{Huber2010FlatBands,Lin2014FlatbandOptical,Peotta2015,PhysRevLett.117.045303,Liang2017BandGeometry,Torma2022,Chau2026QuantumMetricLength}. Although the conventional group velocity vanishes in an exactly flat band, the underlying Bloch wave functions can retain a nontrivial quantum metric and Berry curvature, enabling physical responses that are not captured by ordinary band-dispersion arguments~\cite{Peotta2015,PhysRevLett.117.045303,Liang2017BandGeometry,PhysRevLett.106.236802,Sun2011NearlyFlat,Neupert2011FQH}. The large number of states at the same energy also makes flat-band systems particularly sensitive to interactions and other perturbations~\cite{Mielke1991,Tasaki1992,Huber2010FlatBands,Lin2014FlatbandOptical,PhysRevB.83.220503,Peotta2015,PhysRevLett.117.045303,PhysRevLett.106.236802,Sun2011NearlyFlat,Neupert2011FQH,Regnault2011FCI}. In particular, flat-band superconductors have been proposed as promising platforms for enhanced superconductivity, since the superconducting transition temperature can scale more favorably with interaction strength than in conventional dispersive-band BCS systems, where the critical temperature is exponentially suppressed at weak coupling~\cite{PhysRevB.83.220503,Peotta2015,PhysRevLett.117.045303,Liang2017BandGeometry,Torma2022}.

\begin{figure*}
    \centering
    \includegraphics[width=\textwidth]{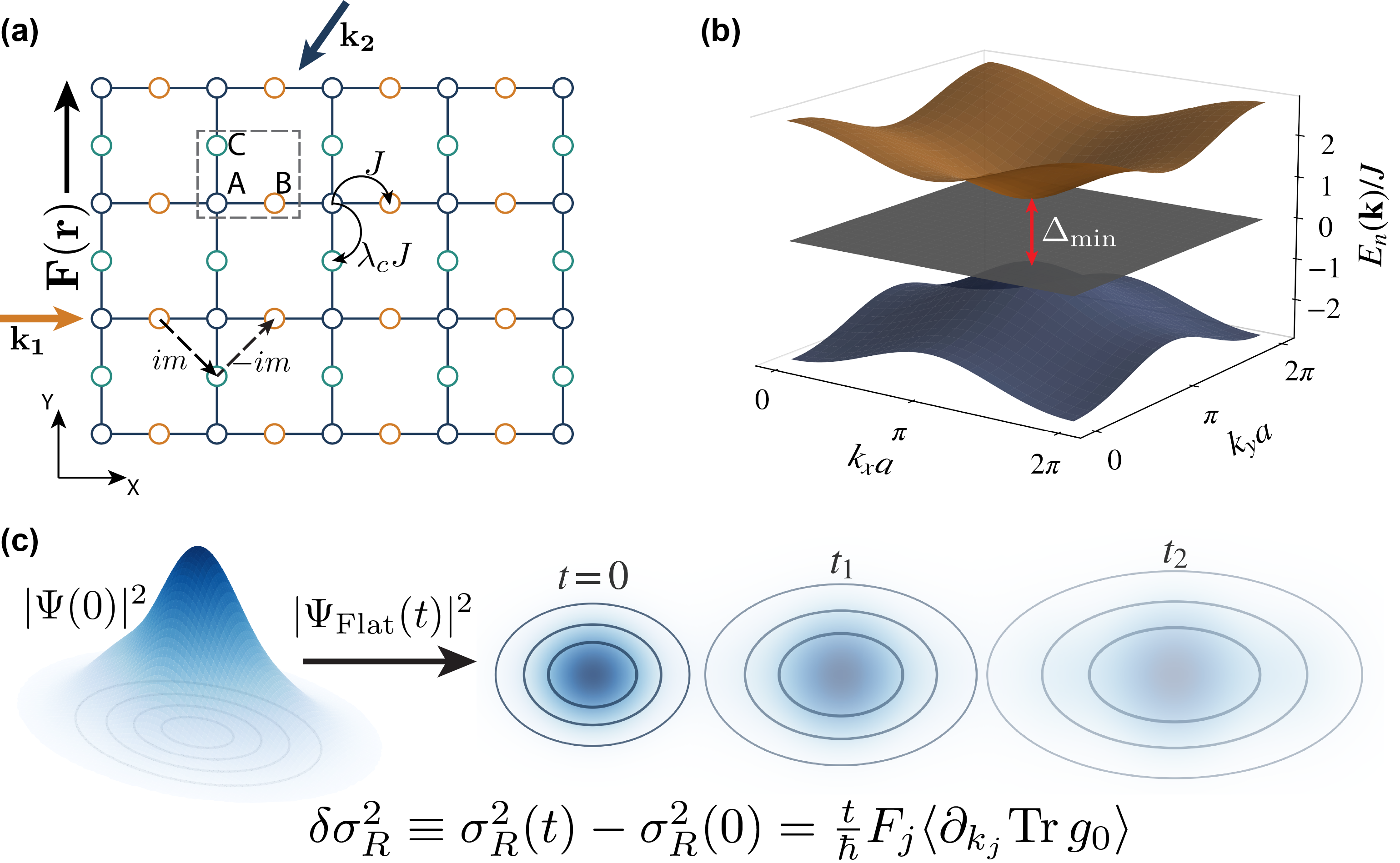}
    \caption{Schematic of the gapped optical Lieb-lattice model and the resulting geometric wave-packet dynamics. (a) Two-dimensional Lieb lattice with sublattices \(A\), \(B\), and \(C\). The \(A\)-\(B\) and \(A\)-\(C\) hopping amplitudes are \(J\) and \(\lambda_c J\), respectively, while the complex \(B\)-\(C\) hoppings \(\pm im\) open a gap while preserving the flat middle band. The Raman wave vectors \(\mathbf{k}_1\) and \(\mathbf{k}_2\) generate the required bond-dependent Peierls phases (see Appendix~\ref{app:raman_implementation}), and \(\mathbf{F}\) denotes the applied weak force. (b) Corresponding band structure, showing the flat band separated from the upper and lower dispersive bands by the minimum gap \(\Delta_{\min}\). (c) Schematic of the force-induced wave-packet spreading. For the flat band considered here, the Berry curvature vanishes and the center of mass remains stationary at leading order, while the real-space variance increases with time. In this limit, the leading change in the total variance is proportional to the wave-packet-averaged momentum-space derivative of the quantum-metric trace, as indicated by the relation shown below the schematic.}
    \label{fig:schematic_lattice}
\end{figure*}

The momentum-space structure of the Bloch eigenstates is characterized by the quantum geometric tensor~\cite{Provost1980,Xiao2010BerryPhase,Yu2025QuantumGeometry,Gao2026QuantumGeometry}, defined for band-$n$ as
\begin{eqnarray}
\mathcal{Q}_n^{ij}(\bm{k})
=
\langle \partial{k_i}u_n|
\left(1-|u_n\rangle\langle u_n|\right)
|\partial_{k_j}u_n\rangle,
\end{eqnarray}
where $\ket{u_n(\bm{k})}$ denotes the cell-periodic Bloch eigenstate. The imaginary part of this tensor determines the Berry curvature, which behaves as an effective magnetic field in momentum space and has long been recognized as a key ingredient in anomalous transport, including the intrinsic anomalous Hall effect and the thermal Hall response~\cite{karplus1954hall,Berry1984,Nagaosa2010AHE,Xiao2010BerryPhase,Qin2011ThermalHall}. The real part defines the quantum metric, $g_n^{ij}(\bm{k})=\mathrm{Re}(\mathcal{Q}_n^{ij}(\bm{k}))$, whose influence on transport is more subtle. Within the conventional semiclassical framework, the quantum metric does not directly generate the anomalous velocity appearing in the equations of motion~\cite{Xiao2010BerryPhase}. Instead, its effects can appear through nonlinear transport responses governed by field induced corrections to the energy and the Berry curvature~\cite{Gao2014PositionalShift,Das2023QuantumMetricNL,Wang2023QuantumMetricTransport,Yu2025QuantumGeometry,Gao2026QuantumGeometry}. This indirect appearance makes it challenging to isolate quantum-metric effects in semiclassical transport measurements and motivates the study of settings where the quantum metric controls the dynamics more directly.

Beyond semiclassical transport, the quantum metric has been shown to play a more direct role in several dynamical and many-body phenomena~\cite{Pruneda2009Wavepacket,PhysRevB.97.201117,Bleu2018QGT,Tan2019QuantumMetric,Gianfrate2020QGT,Peotta2015,PhysRevLett.117.045303,Liang2017BandGeometry,HerzogArbeitman2022Superfluid,Huhtinen2022FlatBand,Zeng2025QMO}. In periodically driven systems, the excitation rate induced by a weak periodic modulation can be directly related to components of the quantum metric, providing an experimentally accessible method for probing the metric of Bloch bands~\cite{PhysRevB.97.201117,Tan2019QuantumMetric}. Related wave-packet studies have shown that quantum geometry can also influence nonadiabatic dynamics and induce oscillatory motion in nearly dispersionless bands~\cite{Pruneda2009Wavepacket,Zeng2025QMO}. These approaches access the quantum metric through excitation rates or dynamical trajectories, providing useful alternatives to transport-based measurements. In flat-band systems, the quantum metric also controls interaction-driven responses in the absence of conventional band dispersion. In particular, the superfluid weight, which characterizes the ability of a system to sustain a dissipationless supercurrent, can acquire a purely geometric contribution determined by the quantum metric of the flat band~\cite{Peotta2015,PhysRevLett.117.045303,Liang2017BandGeometry,HerzogArbeitman2022Superfluid,Huhtinen2022FlatBand,Torma2022}. This result establishes that stable superfluidity and superconductivity can persist even when the conventional contribution associated with band dispersion vanishes. The quantum metric has also been connected to disorder-induced transport in flat-band systems. In disordered one-dimensional Lieb lattices, the quantum metric length, defined through the Brillouin-zone average of the flat-band quantum metric, has been shown to determine relevant length scales of the system, including the localization length in the localized regime and the diffusion length in the diffusive regime~\cite{Chau2026QuantumMetricLength}. These results demonstrate that the quantum metric is not merely a formal geometric property of Bloch wave functions, but can directly govern measurable dynamical, transport, and many-body responses~\cite{Yu2025QuantumGeometry,Gao2026QuantumGeometry}.


Motivated by these developments, we study wave-packet dynamics in clean flat-band systems following the application of a weak external force. We first develop a general description of the dynamics of a wave packet prepared in a well-isolated flat-band. In this limit, the conventional contribution from the band velocity is absent, while the dynamics can still depend on the momentum-space geometry of the flat-band Bloch wave functions~\cite{Zeng2025QMO}. In particular, the motion of the wave-packet center is governed by the Berry curvature, whereas the evolution of its spatial variance contains contributions from both the Berry curvature and the quantum metric~\cite{Xiao2010BerryPhase,Pruneda2009Wavepacket}. This provides a natural way to distinguish the roles of the two components of the quantum geometric tensor in flat-band dynamics and, in the absence of Berry curvature, to isolate the quantum-metric contribution to the wave-packet spreading.


The Lieb lattice is particularly well suited for isolating quantum-metric effects because it combines several properties required for the present proposal. It supports an exactly flat middle band, which can be separated from the neighboring dispersive bands without destroying its flatness; the resulting flat band has vanishing Berry curvature but a nontrivial, momentum-dependent quantum metric; and optical Lieb lattices have already been realized experimentally with controllable tunneling amplitudes and flat-band preparation~\cite{Apaja2010OpticalFlatBand,Goldman2011LiebColdAtoms,doi:10.1126/sciadv.1500854,PhysRevLett.118.175301,taie2020spatial}. This combination removes both the conventional contribution from band dispersion and the leading Berry-curvature contribution to the center-of-mass motion, while retaining a finite quantum-metric response. To isolate the flat band from the neighboring dispersive bands, we introduce complex \(B\)-\(C\) hopping terms that could be implemented using laser-assisted tunneling~\cite{miyake2013realizing,PhysRevLett.111.185301}. These hopping terms open a gap between the flat and dispersive bands while preserving the exact flatness of the middle band and its vanishing Berry curvature. The resulting setup and its force-induced wave-packet dynamics are summarized in Fig.~\ref{fig:schematic_lattice}. In particular, under a weak constant force, the leading short time dynamics within the isolated-flat-band approximation obey
\begin{eqnarray}
\delta\langle\bm r\rangle &=& 0, \\
\delta\sigma_r^2 &=& \frac{t}{\hbar}\bm F\cdot
\left\langle
\bm{\nabla}_{\bm k}\operatorname{Tr}g_0(\bm k)
\right\rangle,
\end{eqnarray}
where \(g_0\) denotes the quantum metric of the flat band, \(\sigma_r^2\) is the total spatial variance of the wave packet, the brackets denote an average over the momentum-space distribution of the initial wave packet, and
\begin{equation}
    \mathrm{Tr}\,g_0(\bm{k})
    =
    \sum_i g_0^{ii}(\bm{k}).
\end{equation} 
is the trace of quantum metric for the flat-band. The change in variance is therefore determined by the wave-packet-averaged momentum-space derivative of the quantum-metric trace along the direction of the applied force. Thus, the wave packet does not move as a whole, but its width can change because of the momentum-space variation of the quantum metric. For comparison, if the cell-periodic Bloch state is independent of momentum, the quantum metric vanishes and this geometric contribution to the spreading is absent. A conventional single-orbital square-lattice band provides a simple example of this limit.

This result provides a possible experimental protocol for probing the quantum metric in clean flat-band systems. Related dynamical approaches have accessed quantum-geometric information through periodic driving, wave-packet motion, and local-quench protocols~\cite{Pruneda2009Wavepacket,PhysRevB.97.201117,Tan2019QuantumMetric,Gianfrate2020QGT,Zeng2025QMO,Komissarov2026FlatBandGeometry}. Periodic-driving protocols relate excitation rates to components of the quantum metric, while previous wave-packet approaches have focused on geometric contributions to the trajectory or oscillatory motion of the wave packet. Here, by contrast, the quantum-metric contribution is extracted from the short time evolution of the real-space wave-packet width under a weak constant force. This does not require semiclassical transport coefficients, disorder-induced length scales, or interaction-driven superfluid responses. By measuring the time dependence of the wave-packet variance, the corresponding wave-packet-averaged momentum-space derivative of the flat-band quantum-metric trace can be obtained. Our analysis therefore identifies wave-packet spreading as a dynamical probe of quantum-metric effects in flat-band systems.

The remainder of the paper is organized as follows. In Sec.~II, we develop the analytical description of force-induced wave-packet dynamics, introduce the gapped Lieb-lattice model, and outline the corresponding experimental protocol. In Sec.~III, we present the numerical results and compare the real-space wave-packet spreading with the analytical quantum-metric prediction. Section~IV summarizes our main results and conclusions. Additional details of the analytical derivation, interband leakage, and the complex \(B\)-\(C\) hopping pattern are provided in the Appendices.

\section{FORCE-INDUCED WAVE-PACKET DYNAMICS IN FLAT BANDS}

We first establish the connection between the real-space dynamics of a flat-band wave packet and the underlying band geometry. We consider an initially localized wave packet confined to an isolated flat band and study its short time response to a weak constant force. This allows the center-of-mass motion to be related to the Berry curvature, while the spatial variance contains contributions from both the Berry curvature and quantum metric.

\subsection{Analytical Framework}

In this section, we derive the dynamics of a general wave packet constructed from the eigenstates of an isolated flat band in a multiband Bloch Hamiltonian. We focus on the short time evolution induced by a constant external force and obtain the corresponding evolution of the first and second moments of the wave packet. Let the initial state be
\begin{eqnarray}
    \ket{\Psi(0)} &=& \int_{\mathrm{BZ}} [d k]\, a_0(\bm{k},0)\ket{\psi_{0\bm{k}}} \nonumber \\
    &=& \int_{\mathrm{BZ}} [d k]\, a_0(\bm{k},0)e^{i\bm{k}\cdot \bm{r}} \ket{u_0(\bm{k})},
    \label{intlwf}
\end{eqnarray}
where $\ket{u_0(\bm{k})}$ is the cell-periodic part of the Bloch eigenstate in the flat band, labeled by $n=0$, and $a_0(\bm{k},0)$ specifies the initial wave-packet profile in momentum space. Here, $[dk]$ denotes the Brillouin-zone integration measure.

We first recall the action of the position operator in the Bloch eigenstate basis. For a multiband system, the position operator satisfies
\begin{eqnarray}
    \hat r_i\ket{\psi_{n\bm{k}}}
    =
    D^i_n\ket{\psi_{n\bm{k}}}
    +
    \sum_{m\neq n}A^i_{mn}(\bm{k})\ket{\psi_{m\bm{k}}},
    \label{roperator}
\end{eqnarray}
where
\begin{eqnarray}
    D^i_n = -i\partial_{k_i} + A^i_n(\bm{k})
\end{eqnarray}
is the band-projected covariant derivative. The first term in Eq.~\eqref{roperator} acts within the subspace of band $n$, while the second term couples the state to the remaining bands through the interband Berry connection $\bm A_{mn}$. The intraband and interband Berry connections are defined as
\begin{eqnarray}
    A^i_n(\bm{k}) = i\braket{u_n(\bm{k})|\partial_{k_i}u_n(\bm{k})}, \label{intraBC}\\
    A^i_{mn}(\bm{k}) = i\braket{u_m(\bm{k})|\partial_{k_i}u_n(\bm{k})}. \label{interBC}
\end{eqnarray}


Upon applying a constant external force, the Hamiltonian acquires the perturbation
\begin{eqnarray}
    \delta H = -\bm{F}\cdot \hat{\bm r}.
\end{eqnarray}
We now evolve the wave packet in Eq.~\eqref{intlwf} under this perturbation. In the isolated flat-band limit, where transitions to the remaining bands are neglected, the time-evolved state can be written as
\begin{eqnarray}
    \ket{\Psi(t)}
    =
    e^{-iE_0t/\hbar}
    \int [dk]\,a_0(\bm{k},t)\ket{\psi_{0\bm{k}}},
    \label{psi(t)}
\end{eqnarray}
where $\ket{\psi_{0\bm{k}}}$ is the Bloch eigenstate of the unperturbed flat band with energy $E_0$. The momentum-space wave-packet amplitude evolves according to
\begin{eqnarray}
    a_0(\bm{k},t)
    =
    e^{\frac{i t}{\hbar}\bm F\cdot \bm D_0}
    a_0(\bm{k},0),
    \label{a0(t)}
\end{eqnarray}
where $\bm D_0$ is the covariant derivative projected onto the flat band. The details of this derivation are given in Appendix~\ref{appendix: Appendix A}.

The validity of Eq.~\eqref{psi(t)} requires the force-induced coupling to the dispersive bands to remain small compared with the corresponding band gaps. A sufficient perturbative condition is
\begin{eqnarray}
    \max_{\bm{k}, n}\frac{\left|\bm F\cdot \bm A_{n0}(\bm{k})\right|}
    {\left|\Delta_{n0}(\bm{k})\right|}
    \ll 1,
    \label{bandgap}
\end{eqnarray}
for all relevant momenta $\bm{k}$ and bands $n\neq 0$, where $\left|\Delta_{n0}(\bm{k})\right|$ is the energy separation between the flat band and band $n$. A band-independent sufficient criterion can be expressed in terms of manifestly gauge invariant quantities as
\begin{eqnarray}
    \max_{\bm{k}}
    \frac{\sqrt{F_iF_jg^{ij}_0(\bm{k})}}
    {\Delta_{\mathrm{min}}}
    \ll 1,
    \label{bandgap_metric}
\end{eqnarray}
where $\Delta_{\mathrm{min}}$ is the minimum gap between the flat band and the remaining bands over the relevant region of the Brillouin zone, and $g_n^{ij}$ is the quantum metric of band n defined by
\begin{eqnarray}
    g^{ij}_n = \mathrm{Re}(\braket{\partial_in|\partial_jn}-A^i_nA^j_n),
\end{eqnarray}
where $\mathrm{Re}(z)$ denotes the real part of $z$, and $A^i_n$ is the intraband Berry connection corresponding to band with index $n$. Under this condition, interband leakage is perturbatively suppressed, and the dynamics may be described by the projected flat-band evolution in Eq.~\eqref{psi(t)}. We shall use this approximation throughout the analytical treatment below.

To characterize the wave-packet dynamics, we evaluate the time dependence of its position moments. In particular, we consider the first moment, which describes the center-of-mass motion, and the second central moment, which characterizes the spatial spread of the wave packet. We focus on the short time regime defined by
\begin{eqnarray}
    \frac{|\bm F|t\sigma_r}{\hbar} \ll 1,
    \label{smalltime}
\end{eqnarray}
where $\sigma_r$ denotes the initial real-space standard deviation of the wave packet. In this regime, Eq.~\eqref{a0(t)} can be expanded as
\begin{eqnarray}
    a_0(\bm{k},t)
    \approx
    \left(1+\frac{i t}{\hbar}\bm F\cdot \bm D_0\right)
    a_0(\bm{k},0).
    \label{a0(t)small}
\end{eqnarray}

\begin{widetext}
Substituting Eq.~\eqref{a0(t)small} into Eq.~\eqref{psi(t)}, we obtain the leading change in the first moment and in the wave-packet variance,
\begin{eqnarray}
    \delta\langle r_i\rangle
    &\equiv&
    \langle r_i(t)\rangle-\langle r_i(0)\rangle
    =
    -\frac{t}{\hbar}F_j\langle \Omega_{ij}\rangle,
    \label{deltar}
    \\
    \delta \sigma_{r_i}^2
    &\equiv&
    \sigma_{r_i}^2(t)-\sigma_{r_i}^2(0)
    =
    \frac{t}{\hbar}F_j
    \bigg[
    -\langle \{D_0^i,\Omega_{ij}\}\rangle
    +
    2\langle D_0^i\rangle\langle \Omega_{ij}\rangle
    +
    \langle \partial_{k_j}g_0^{ii}\rangle
    \bigg].
    \label{deltarsq}
\end{eqnarray}
\end{widetext}

Equations~\eqref{deltar} and \eqref{deltarsq} show that, in the projected flat-band dynamics, the center-of-mass motion is governed by the Berry curvature, whereas the wave-packet spread contains contributions from both the Berry curvature and the quantum metric. These results illustrate how the quantum geometry of the flat band influences wave-packet dynamics even when the conventional group velocity vanishes. In the flat-band Lieb lattice model considered below, the Berry curvature of the flat band vanishes identically. Equations~\eqref{deltar} and \eqref{deltarsq} then predict that the wave-packet center of mass remains stationary, while the leading change in the spread is controlled by the wave-packet-averaged momentum derivative of the relevant quantum-metric component.

\subsection{Lieb-Lattice Model}
\label{sec:lieb_lattice_model}

The analytical result derived above is most cleanly isolated in a system with an exactly flat and well-separated band, vanishing Berry curvature, and a nontrivial momentum-dependent quantum metric. We therefore specialize to the gapped Lieb lattice introduced in Fig.~\ref{fig:schematic_lattice}. The Lieb geometry provides an exactly flat middle band, while the additional complex \(B\)-\(C\) hopping introduced below separates this band from the neighboring dispersive bands without destroying its flatness or generating a finite Berry curvature. This combination allows the quantum-metric contribution to the wave-packet dynamics to be studied without competing contributions from conventional band dispersion or leading-order Berry-curvature-induced center-of-mass motion.

Each unit cell contains three inequivalent sites, denoted by \(A\), \(B\), and \(C\), with relative intracell positions
\begin{equation}
    \boldsymbol{\tau}_A=(0,0),\qquad
    \boldsymbol{\tau}_B=
    \left(\frac{a}{2},0\right),\qquad
    \boldsymbol{\tau}_C=
    \left(0,\frac{a}{2}\right),
    \label{eq:lieb_basis_positions}
\end{equation}
where \(a\) is the lattice constant.

Neglecting interactions, the single-particle tight-binding Hamiltonian is written as
\begin{equation}
    H_0=H_{AB}+H_{AC}+H_{BC}.
    \label{eq:real_space_hamiltonian_decomposition}
\end{equation}
The nearest-neighbor tunneling terms are
\begin{align}
    H_{AB}
    &=
    J\sum_{\mathbf R}
    \left(
        a_{\mathbf R}^{\dagger}b_{\mathbf R}
        +
        a_{\mathbf R}^{\dagger}b_{\mathbf R-a\hat{x}}
    \right)
    +\mathrm{H.c.},
    \label{eq:real_space_AB}
    \\
    H_{AC}
    &=
    \lambda_c J\sum_{\mathbf R}
    \left(
        a_{\mathbf R}^{\dagger}c_{\mathbf R}
        +
        a_{\mathbf R}^{\dagger}c_{\mathbf R-a\hat{y}}
    \right)
    +\mathrm{H.c.},
    \label{eq:real_space_AC}
\end{align}
where \(a_{\mathbf R}\), \(b_{\mathbf R}\), and \(c_{\mathbf R}\) annihilate an atom on the corresponding sublattice in unit cell \(\mathbf R\). The tunneling amplitude \(J\) sets the energy scale, while \(\lambda_c\) controls the tunneling along the \(A\)-\(C\) bonds relative to the \(A\)-\(B\) bonds.

The flat band is separated from the dispersive bands by introducing complex tunneling between the \(B\) and \(C\) sublattices. The required real-space term is
\begin{align}
    H_{BC}
    =
    im\sum_{\mathbf R}
    \Big(
        &b_{\mathbf R}^{\dagger}c_{\mathbf R}
        -
        b_{\mathbf R}^{\dagger}c_{\mathbf R-a\hat{y}}
        -
        b_{\mathbf R}^{\dagger}c_{\mathbf R+a\hat{x}}
        \nonumber\\
        &+
        b_{\mathbf R}^{\dagger}
        c_{\mathbf R+\hat{x}-a\hat{y}}
    \Big)
    +\mathrm{H.c.},
    \label{eq:real_space_BC}
\end{align}
where \(m\) controls the gap opened at the band-touching point. Thus, the four diagonal \(B\)-\(C\) tunneling processes have amplitudes
\begin{equation}
    \left\{
    +im,-im,-im,+im
    \right\}.
    \label{eq:BC_phase_pattern}
\end{equation}
The real-space origin of Eq.~\eqref{eq:momentum_BC} and the corresponding bond-dependent phase pattern are discussed in Appendix~\ref{app:raman_implementation}.

Using the physical positions of the three sites, we define
\begin{equation}
    \alpha_{\mathbf R}
    =
    \frac{1}{\sqrt{N_{\mathrm{uc}}}}
    \sum_{\mathbf k}
    e^{i\mathbf k\cdot
    (\mathbf R+\boldsymbol{\tau}_{\alpha})}
    \alpha_{\mathbf k},
    \qquad
    \alpha=A,B,C,
    \label{eq:fourier_sublattice}
\end{equation}
where \(N_{\mathrm{uc}}\) is the number of unit cells. Equations~\eqref{eq:real_space_AB} and \eqref{eq:real_space_AC} then give
\begin{align}
    H_{AB}(\mathbf k)
    &=
    2J\cos\left(\frac{k_xa}{2}\right),
    \\
    H_{AC}(\mathbf k)
    &=
    2\lambda_cJ
    \cos\left(\frac{k_ya}{2}\right).
\end{align}
The complex tunneling pattern in Eq.~\eqref{eq:real_space_BC} similarly gives
\begin{equation}
    H_{BC}(\mathbf k)
    =
    4im
    \sin\left(\frac{k_xa}{2}\right)
    \sin\left(\frac{k_ya}{2}\right).
    \label{eq:momentum_BC}
\end{equation}
In the sublattice basis \(\Psi_{\mathbf k}= (a_{\mathbf k},b_{\mathbf k},c_{\mathbf k})^{T}\), the resulting Bloch Hamiltonian is therefore

\begin{widetext}
\begin{equation}
    H_0(\mathbf k)
    =
    \begin{pmatrix}
        0 &
        2J\cos(k_xa/2) &
        2\lambda_cJ\cos(k_ya/2)
        \\
        2J\cos(k_xa/2) &
        0 &
        4im\sin(k_xa/2)\sin(k_ya/2)
        \\
        2\lambda_cJ\cos(k_ya/2) &
        -4im\sin(k_xa/2)\sin(k_ya/2) &
        0
    \end{pmatrix}.
    \label{eq:cold_atom_lieb_hamiltonian}
\end{equation}
\end{widetext}

\begin{figure*}
    \centering
    \includegraphics[width=\textwidth]{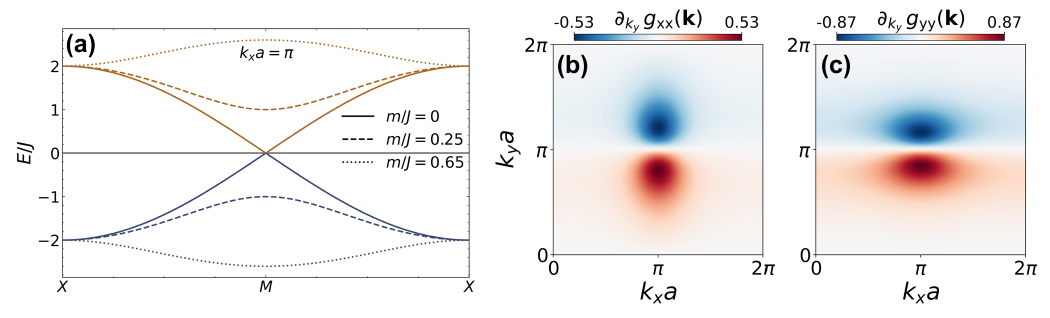}
    \caption{Single-particle properties of the gapped Lieb lattice. (a) Band structure showing the dispersionless flat band separated from the upper and lower dispersive bands by the minimum gap \(\Delta_{\min}\). (b,c) Momentum-space distributions of \(\partial_{k_y}g_{xx}(\mathbf{k})\) and \(\partial_{k_y}g_{yy}(\mathbf{k})\), respectively.}
    \label{fig:lieb_features}
\end{figure*}

The complex \(B\)-\(C\) tunneling opens a gap between the flat band and the two dispersive bands while preserving the exactly dispersionless middle band. The resulting single-particle band structure and the momentum dependence of the relevant quantum-metric components are shown in Fig.~\ref{fig:lieb_features}.

For the theoretical construction used in the numerical calculations, we consider a localized Gaussian wave packet centered at the quasimomentum \(\mathbf{k}_0\). For a sufficiently narrow momentum-space distribution, the flat-band eigenvector varies only weakly over the occupied momenta. We therefore construct the initial wave packet using the flat-band spinor evaluated at the central quasimomentum,
\begin{equation}
    \eta
    =
    u_0(\mathbf{k}_0)
    =
    \frac{1}{\mathcal N_u}
    \begin{pmatrix}
        -iM_0\\
        -B_0\\
        A_0
    \end{pmatrix},
\end{equation}
where
\begin{align}
    M_0
    &=
    4m
    \sin\!\left(\frac{k_{0x}a}{2}\right)
    \sin\!\left(\frac{k_{0y}a}{2}\right),\\
    A_0
    &=
    2J
    \cos\!\left(\frac{k_{0x}a}{2}\right),\\
    B_0
    &=
    2\lambda_c J
    \cos\!\left(\frac{k_{0y}a}{2}\right),
\end{align}
and \(\mathcal N_u\) is the normalization constant. The corresponding single-particle state takes the form
\begin{equation}
    |\Psi_G\rangle
    =
    \frac{1}{\mathcal N}
    \sum_{\mathbf R,\alpha}
    e^{i\mathbf k_0\cdot\mathbf r_{\mathbf R\alpha}}
    e^{
    -|\mathbf r_{\mathbf R\alpha}-\mathbf r_0|^2/
    (2\sigma_r^2)
    }
    \eta_{\alpha}
    |\mathbf R,\alpha\rangle,
    \label{eq:cold_atom_gaussian}
\end{equation}
where \(\sigma_r\) determines the real-space extent of the wave packet.

Although the initial state is constructed using the flat-band eigenvector at the central quasimomentum, the finite momentum width of the Gaussian wave packet samples neighboring crystal momenta for which the Bloch eigenvector varies continuously, \(u_0(\mathbf{k})\neq u_0(\mathbf{k}_0)\). Consequently, the state in Eq.~\eqref{eq:cold_atom_gaussian} is not an exact superposition of flat-band Bloch eigenstates and contains a small contribution from the dispersive bands. For a sufficiently narrow momentum distribution, this contribution remains small and the overlap with the flat-band manifold correspondingly remains close to unity. To isolate the intrinsic flat-band dynamics described by the analytical theory in our numerical calculations, we project the constructed wave packet onto the flat-band subspace prior to time evolution,
\begin{equation}
    |\Psi(0)\rangle
    =
    \frac{P_0|\Psi_G\rangle}
    {\sqrt{\langle\Psi_G|P_0|\Psi_G\rangle}},
    \label{eq:experimental_flat_loading}
\end{equation}
where \(P_0\) projects onto the flat-band manifold. The corresponding projection weight,
\begin{equation}
    W_0
    =
    \langle\Psi_G|P_0|\Psi_G\rangle,
\end{equation}
quantifies the overlap of the constructed Gaussian state with the exact flat-band subspace and, for the parameters considered in this work, is found to be close to unity.

\subsection{Experimental Protocol}
\label{sec:cold_atom_protocol}

We now outline a possible cold-atom protocol for realizing the model and measuring the predicted wave-packet spreading. The proposal requires four main ingredients: realization of the gapped optical Lieb lattice, preparation of a localized wave packet with substantial flat-band occupation at a chosen central quasimomentum, application of a weak uniform force, and measurement of the resulting real-space density profile and variance.

Optical Lieb lattices with controllable sublattice potentials and tunneling amplitudes have been proposed and experimentally realized using superposed optical standing waves~\cite{Apaja2010OpticalFlatBand,Goldman2011LiebColdAtoms,doi:10.1126/sciadv.1500854,taie2020spatial}. The relative tunneling amplitude \(\lambda_c\) can be controlled by varying the corresponding lattice barriers~\cite{taie2020spatial} or the strengths of laser-assisted tunneling processes~\cite{PhysRevLett.107.255301,PhysRevLett.111.185301,PhysRevLett.111.185302}. The additional bond-dependent complex \(B\)-\(C\) tunneling required to isolate the flat band could be engineered by adapting established Raman- and laser-assisted tunneling schemes, in which an energy offset suppresses the bare tunneling and a resonant Raman process restores it with a controllable Peierls phase. These techniques restore otherwise suppressed tunneling while imprinting controllable Peierls phases on the corresponding matrix elements~\cite{PhysRevLett.107.255301,PhysRevLett.111.185301,PhysRevLett.111.185302}.

The preparation of the initial state is motivated by experimentally established techniques for producing localized matter-wave packets in optical lattices and transferring atoms into the flat band of an optical Lieb lattice~\cite{pedersen2013production,Kowalski2025SpatioTemporal,doi:10.1126/sciadv.1500854}. Localized matter-wave packets have been produced and manipulated in combined optical-lattice and confining potentials~\cite{pedersen2013production}, while coherent population transfer into the flat band of an optical Lieb lattice has been demonstrated using phase imprinting~\cite{doi:10.1126/sciadv.1500854}. Spatial adiabatic passage provides an experimentally demonstrated method for preparing and following the flat-band dark state by controlling the sublattice tunneling amplitudes~\cite{taie2020spatial}.


The central quasimomentum \(\bm k_0\) can be controlled experimentally by applying a uniform force for a prescribed time. Controlled quasimomentum transport using Bloch oscillations has been demonstrated in optical lattices~\cite{atala2013direct}, and transport of a Bose-Einstein condensate to a desired quasimomentum by application of a constant force has been demonstrated directly in an optical Lieb lattice~\cite{PhysRevLett.118.175301}.

Consistent with the single-particle description used in this work, interactions should be sufficiently weak that interaction-induced spreading remains negligible over the measurement interval. For atomic species with an accessible Feshbach resonance, the scattering length may be tuned close to a zero crossing to further suppress interactions during the dynamical protocol~\cite{Chin2010Feshbach,Volz2003ZeroCrossing}.


After the flat-band wave packet is prepared at the desired central quasimomentum, a uniform force is switched on along the \(i=x\) or \(y\) direction. For neutral atoms, such a force can be generated by accelerating the optical lattice, which produces a controlled inertial force in the lattice frame~\cite{Madison1999AcceleratingLattice,Li2016WilsonLines}. Constant-force transport through momentum space has also been demonstrated directly in an optical Lieb lattice~\cite{PhysRevLett.118.175301}. In the lattice frame, the acceleration produces an inertial potential
\begin{equation}
    V_F=-F_i\hat r_i,
    \label{eq:atomic_force}
\end{equation}
where
\begin{equation}
    F_i=M a_i
\end{equation}
in magnitude, with \(M\) the atomic mass and \(a_i\) the lattice acceleration.


The post-quench Hamiltonian is therefore
\begin{equation}
    H_F=H_0-F_i\hat r_i.
    \label{eq:cold_atom_postquench}
\end{equation}
The applied force changes the atomic quasimomentum according to the acceleration theorem,
\begin{equation}
    \hbar\dot{k}_i=F_i.
    \label{eq:acceleration_theorem}
\end{equation}
For a constant force,
\begin{equation}
    k_i(t)=k_i(0)+\frac{F_i t}{\hbar}.
    \label{acceleratio_theorem}
\end{equation}

\begin{figure*}
    \centering
    \includegraphics[width=\textwidth]{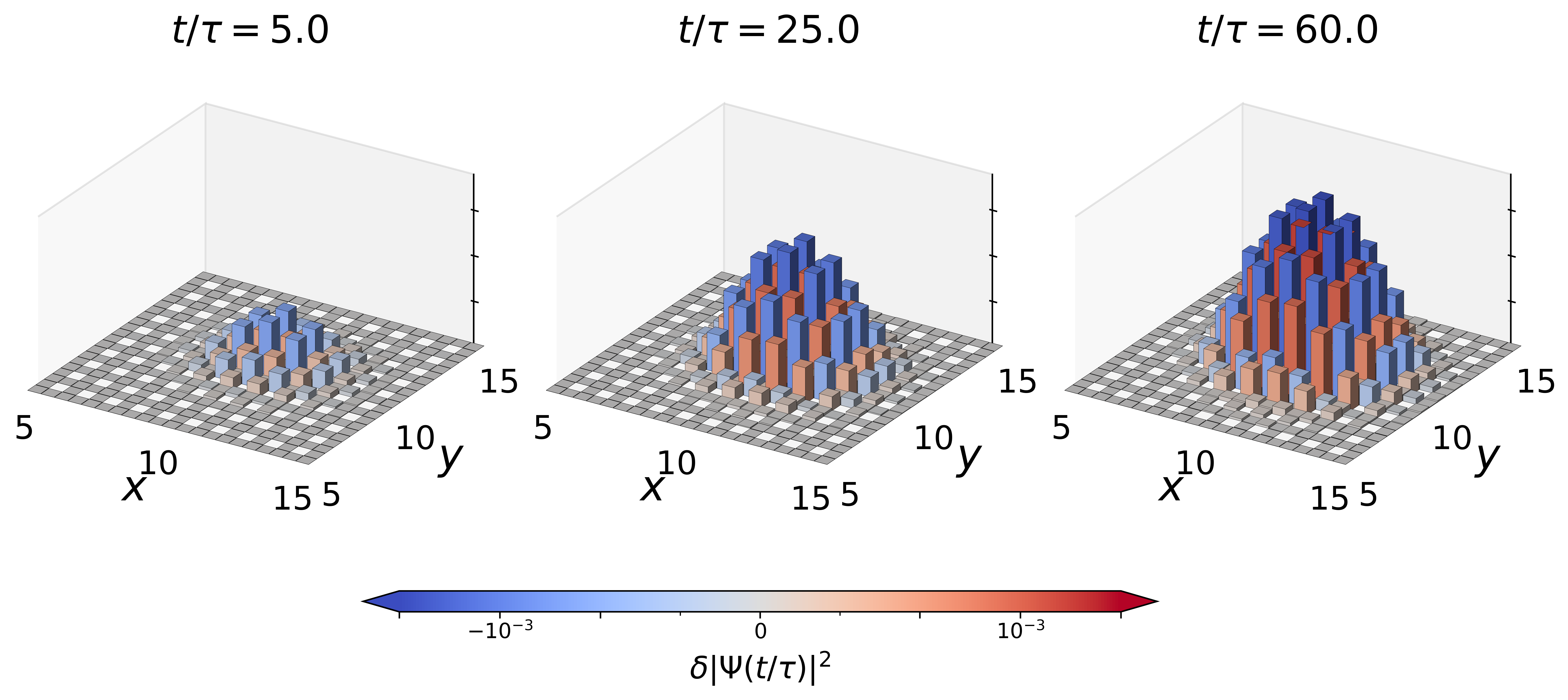}
    \caption{Change in probability density distribution associated with the wave packet, \(\delta\rho_{\mathrm{flat}}(\mathbf r,t)=|\Psi_{\mathrm{flat}}(\mathbf r,t)|^2-|\Psi_{\mathrm{flat}}(\mathbf r,0)|^2\), at \(t/\tau=5,\ 25,\) and \(60\) following the application of a uniform force. The wave packet develops a bimodal profile with increasing real-space spread. Parameters are \(L_x=L_y=21\), \(m/J=0.25\), \(Fa/J=0.005\), \(\mathbf{k}_0a=(\pi/2,\pi/2)\), and \(\sigma_r/a=1.5\), with periodic boundary conditions.}
    \label{fig:PB_density_snapshot}
\end{figure*}


The resulting dynamics can be measured from the real-space atomic density. Real-space relaxation of an initially localized atomic distribution in an optical lattice has been measured directly by preparing a spatially localized spin-tagged population and following its subsequent density evolution~\cite{Kowalski2025SpatioTemporal}. Established in situ fluorescence-imaging techniques also allow the spatial distribution of ultracold atoms in optical lattices to be reconstructed with single-site resolution~\cite{bakr2009quantum,sherson2010single,cheuk2015quantum,parsons2015site}. Such measurements can be used to extract spatial density and variance profiles directly from the recorded atomic distribution~\cite{sherson2010single}. Repeating the preparation and evolution for different values of \(t\) therefore reconstructs the time-dependent density profile of the cloud. As an initial illustration of the force-induced dynamics, Fig.~\ref{fig:PB_density_snapshot} shows the change in the probability density distribution associated with the wave-packet wavefunction at representative times after the force is applied. The snapshots provide a direct visualization of the redistribution of the initially localized wave packet that is quantified below through its real-space variance.


For a site-resolved measurement, let \(n_{\mathbf R\alpha}(t)\) denote the detected occupation of sublattice site \(\alpha\) in unit cell \(\mathbf R\). The total number of detected atoms is
\begin{equation}
    N_{\mathrm{at}}
    =
    \sum_{\mathbf R,\alpha}
    n_{\mathbf R\alpha}(t),
\end{equation}
and the center of mass is
\begin{equation}
    \overline{r}_i(t)
    =
    \frac{1}{N_{\mathrm{at}}}
    \sum_{\mathbf R,\alpha}
    r_{\mathbf R\alpha,i}
    n_{\mathbf R\alpha}(t).
    \label{eq:measured_com}
\end{equation}
The directional variance is obtained from the same density image,
\begin{equation}
    \sigma_i^2(t)
    =
    \frac{1}{N_{\mathrm{at}}}
    \sum_{\mathbf R,\alpha}
    \left[
        r_{\mathbf R\alpha,i}
        -
        \overline{r}_i(t)
    \right]^2
    n_{\mathbf R\alpha}(t).
    \label{eq:measured_variance}
\end{equation}
The total width and its change are then
\begin{align}
    \sigma_r^2(t)
    &=
    \sigma_x^2(t)+\sigma_y^2(t),
    \\
    \delta\sigma_r^2(t)
    &=
    \sigma_r^2(t)-\sigma_r^2(0).
\end{align}

Within the short time and isolated-band regimes, the initial slope of the measured variance is predicted to satisfy
\begin{equation}
    \left.
    \frac{d\delta\sigma_r^2}{dt}
    \right|_{t\rightarrow0}
    =
    \frac{F_i}{\hbar}
    \left\langle
        \partial_{k_i}
        \operatorname{Tr}g_0(\mathbf k)
    \right\rangle.
    \label{eq:experimental_metric_extraction}
\end{equation}
The proposed measurement therefore does not require reconstruction of the full Bloch wave function. Instead, the wave-packet averaged momentum-space derivative of the flat-band quantum-metric trace is inferred from the force-induced growth of the real-space atomic cloud width.


The protocol must remain within the same short time and weak-interband coupling regime assumed in the analytical treatment. In addition to keeping interaction-induced spreading negligible over the measured time interval, the leading-order result requires
\begin{equation}
    \frac{|F_i|\sigma_r t}{\hbar}\ll1,
    \label{eq:cold_atom_short_time}
\end{equation}
while interband transitions are suppressed when
\begin{equation}
    \max_{\mathbf k}
    \frac{
        |F_i|\sqrt{g_{0}^{ii}(\mathbf k)}
    }{
        \Delta_{\min}
    }
    \ll1.
    \label{eq:cold_atom_isolated_band}
\end{equation}
Here, \(\Delta_{\min}\) is the minimum energy separation between the flat band and the nearest dispersive band over the momentum-space region sampled by the wave packet. The numerical leakage analysis presented below determines the extent to which the latter condition is satisfied for the parameters used in the dynamical simulations.

\section{Numerical Results}
\label{sec:numerical_results}

We numerically investigate the real-space dynamics of Gaussian wave packets in the gapped Lieb lattice and compare the resulting spreading with the analytical prediction derived above. Throughout this section, lengths are measured in units of the lattice constant \(a\), energies in units of the nearest-neighbor tunneling amplitude \(J\), and time in units of \(\tau=\hbar/J\). The force, complex next-nearest-neighbor hopping, and wave-packet spreading are therefore expressed as \(Fa/J\), \(m/J\), and \(\delta\sigma_R^2/a^2\), respectively.

\subsection{Flat-band wave-packet dynamics}

The numerical simulations are performed on the gapped Lieb lattice introduced in Sec.~\ref{sec:cold_atom_protocol}. The single-particle spectrum and the momentum-space gradients of the flat-band quantum metric are shown in Fig.~\ref{fig:lieb_features}. The complex \(B\)-\(C\) hopping \(m\) opens a finite gap \(\Delta_{\min}\) separating the dispersionless flat band from the upper and lower dispersive bands. For sufficiently weak applied forces, this gap suppresses Landau-Zener transitions to the dispersive bands~\cite{morsch2006dynamics,Zenesini2009LandauZener}, allowing the dynamics to remain predominantly within the flat-band manifold.

For the flat band considered here, the leading short time change in the wave-packet spread is
\begin{equation}
    \frac{\delta\sigma_R^2}{a^2}
    =
    \frac{F_i a}{J}
    \frac{t}{\tau}
    \left\langle
    \partial_{k_i}
    \operatorname{Tr} g_0(\mathbf{k})
    \right\rangle,
    \label{eq:linear_spreading_prediction}
\end{equation}
where \(F_i\) is the applied force along \(i=x,y\). Figures~\ref{fig:lieb_features}(b) and (c) show the corresponding distributions of \(\partial_{k_y}g_{xx}\) and \(\partial_{k_y}g_{yy}\).

Because the initial wave packet occupies a finite region of momentum space, the response is determined by the wave-packet average of the metric derivative rather than its value at \(\mathbf{k}_0\) alone. We obtain the normalized momentum-space weight
\begin{equation}
    w(\mathbf{k})
    =
    \sum_{\alpha}
    |\Psi_{\alpha}(\mathbf{k},0)|^2,
    \qquad
    \sum_{\mathbf{k}}
    w(\mathbf{k})=1,
    \label{eq:momentum_weight}
\end{equation}
and evaluate
\begin{equation}
    \left\langle
    \partial_{k_i}
    \operatorname{Tr}g_0
    \right\rangle
    =
    \sum_{\mathbf{k}}
    w(\mathbf{k})
    \partial_{k_i}
    \operatorname{Tr}g_0(\mathbf{k}),
    \label{eq:weighted_metric_derivative}
\end{equation}
with the momentum derivatives evaluated on the periodic Brillouin-zone grid using centered finite differences. The leading-order prediction is expected to hold within the short time regime
\begin{equation}
    \frac{t}{\tau}
    \ll
    \frac{1}
    {\sigma(F_i a/J)},
    \label{eq:short_time_condition}
\end{equation}
where \(\sigma=\sigma_r/a\).


For the numerical evolution, we use the Gaussian wave packet introduced in Sec.~\ref{sec:cold_atom_protocol}, centered at \(\mathbf{k}_0\), and project it onto the flat-band manifold of the force-free Hamiltonian \(H_0\). For a finite lattice containing \(L_x\times L_y\) unit cells, the corresponding projector is
\begin{equation}
    P_0
    =
    \sum_{\nu\in\mathrm{flat}}
    |\phi_\nu^{(0)}\rangle
    \langle\phi_\nu^{(0)}|,
    \label{eq:flatband_projector}
\end{equation}
and the initial state used for the time evolution is
\begin{equation}
    |\Psi(0)\rangle
    =
    \frac{P_0|\Psi_G\rangle}
    {\sqrt{\langle\Psi_G|P_0|\Psi_G\rangle}}.
    \label{eq:projected_initial_state}
\end{equation}
The unprojected Gaussian already has a flat-band weight
\(W_0=\langle\Psi_G|P_0|\Psi_G\rangle\) close to unity for the parameter regimes considered here.

%
%
%


At \(t=0\), a uniform force is applied along either the \(x\) or \(y\) direction, giving
\begin{equation}
    H_F
    =
    H_0
    -
    F_i\hat r_i.
    \label{eq:post_quench_hamiltonian}
\end{equation}
The projected initial state is then evolved under the full post-quench Hamiltonian \(H_F\).

%

To visualize the redistribution within the flat-band manifold, we additionally project the evolved state onto the flat band,
\begin{equation}
    |\Psi_{\mathrm{flat}}(t)\rangle
    =
    \frac{
        P_0|\Psi(t)\rangle
    }{
        \sqrt{
        \langle\Psi(t)|P_0|\Psi(t)\rangle
        }
    }.
    \label{eq:normalized_time_projected_state}
\end{equation}
Figure~\ref{fig:PB_density_snapshot} shows the evolution of the probability density distribution density at \(t/\tau=5\), \(25\), and \(60\). Since the plotted quantity is the density difference \(\delta\rho_{\mathrm{flat}}(\mathbf r,t)\), the red regions indicate an increase in probability relative to the initial state, while the blue regions indicate a corresponding depletion. The spatial pattern therefore reflects both the spreading of the initially localized wave packet and a redistribution of probability among the different sublattice sites of the Lieb lattice. As the evolution proceeds, regions of increased and decreased weight extend over a progressively larger area and the dominant features move farther from the initial wave-packet center, leading to an increase in the total real-space variance.

\begin{figure}
    \centering
    \includegraphics[width=\linewidth]{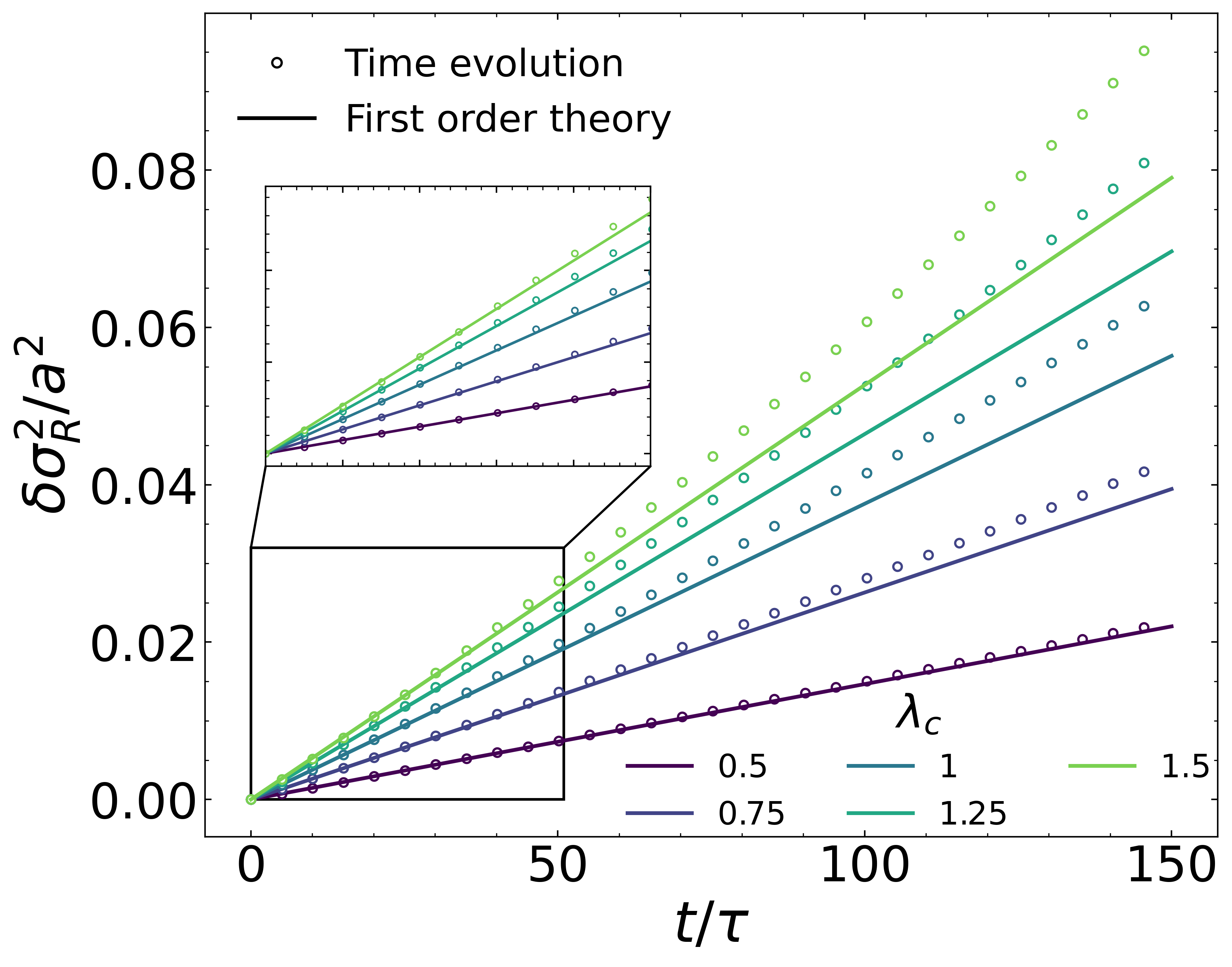}
    \caption{Wave-packet spreading, \(\delta\sigma_R^2/a^2\), as a function of \(t/\tau\) for different values of \(\lambda_c\). Circular markers show the numerical time evolution, while solid lines show the leading order geometric prediction from Eq.~\eqref{eq:linear_spreading_prediction}.}
    \label{fig:WP_spreading_lambda}
\end{figure}

\subsection{Quantum-metric induced wave-packet dynamics}

We characterize the spreading of the full time-evolved state through
\begin{equation}
    \sigma_i^2(t)
    =
    \langle\hat{r}_i^2\rangle_t
    -
    \langle\hat{r}_i\rangle_t^2,
    \qquad i=x,y,
    \label{eq:directional_variance}
\end{equation}
with the total variance
\begin{equation}
    \sigma_R^2(t)
    =
    \sigma_x^2(t)+\sigma_y^2(t),
    \label{eq:total_variance}
\end{equation}
and
\begin{equation}
    \frac{\delta\sigma_R^2(t)}{a^2}
    =
    \frac{\sigma_R^2(t)-\sigma_R^2(0)}{a^2}.
    \label{eq:delta_variance}
\end{equation}


Figure~\ref{fig:WP_spreading_lambda} compares the numerical spreading with the geometric prediction in Eq.~\eqref{eq:linear_spreading_prediction} for several values of \(\lambda_c\). Varying \(\lambda_c\) changes the flat-band Bloch eigenstates and therefore the momentum dependence of the quantum metric. The numerical evolution follows the corresponding linear prediction throughout the short time regime in Eq.~\eqref{eq:short_time_condition}.

\begin{figure}
    \centering
    \includegraphics[width=\linewidth]{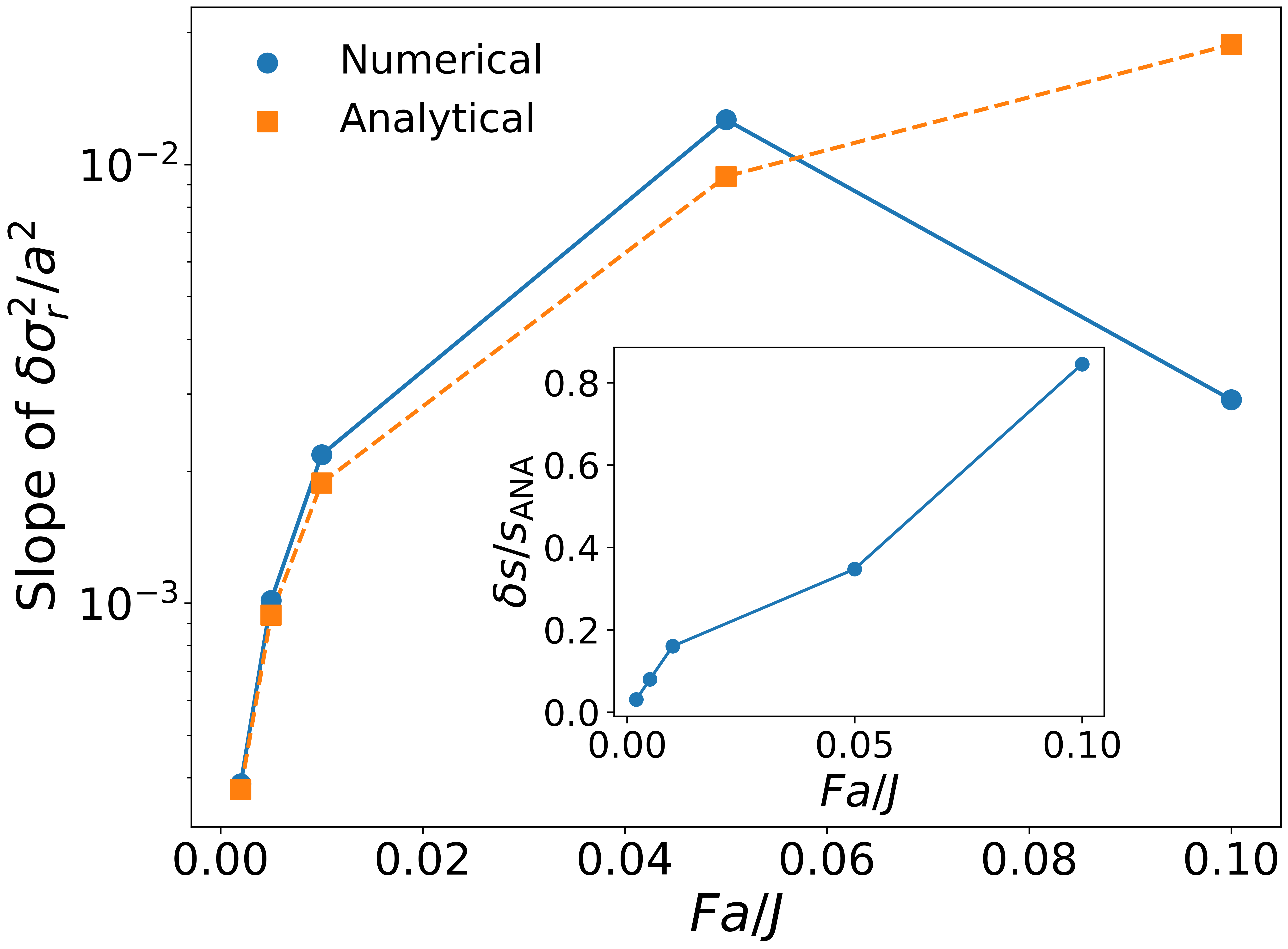}
    \caption{Comparison of the numerically extracted and analytically predicted linear spreading coefficients for a force applied along the \(y\) direction. Numerical slopes are obtained from fits over \(0\leq t\leq t_{\max}\), with \(t_{\max}/\tau=1/[4\sigma(Fa/J)]\). The inset shows the relative deviation, \(\delta s/s_{\mathrm{ANA}}\), where \(\delta s=\lvert s_{\mathrm{NUM}}-s_{\mathrm{ANA}}\rvert\).}
    \label{fig:Slope_comparison}
\end{figure}

A more quantitative comparison is shown in Fig.~\ref{fig:Slope_comparison}, where the numerical linear spreading coefficient is compared directly with the analytical prediction as the applied force is varied. The numerical coefficient is extracted from a linear fit over \(0\leq t\leq t_{\max}\), with \(t_{\max}/\tau=1/[4\sigma(Fa/J)]\). The agreement remains strong in the weak-force regime, while the deviation increases with force as higher-order corrections and interband mixing become more important.

The agreement in both the time-dependent spreading and the extracted linear coefficient shows that the short time dynamics are governed by the momentum-space variation of the flat-band quantum metric. Deviations outside this regime arise from higher-order dynamics and interband mixing; their contribution is examined further in Appendix~\ref{appendix: Leakage}.

\section{Conclusions}

In this work, we have shown that the real-space dynamics of a wave packet in an isolated flat band provide access to the underlying quantum geometry. Starting from a general multiband description, we derived the short time evolution of the first and second moments of a flat-band wave packet under a weak constant force. While the center-of-mass motion is governed by the Berry curvature, the wave-packet spread contains an additional contribution determined by the momentum-space variation of the quantum metric. This geometric contribution remains finite even in the absence of conventional band dispersion.

We illustrated this result using a gapped Lieb lattice with an exactly flat middle band, vanishing Berry curvature, and a nontrivial quantum metric. In this case, the center of mass remains stationary at leading order, while the change in the total real-space variance is linear in time and proportional to the wave-packet-averaged momentum-space derivative of the quantum-metric trace along the direction of the applied force. Exact real-space time evolution confirms this prediction throughout the expected short time and weak-force regime. By varying the lattice anisotropy and the applied force, we find that the numerical spreading rate follows the corresponding geometric prediction, while deviations at larger forces and longer times arise from higher-order corrections and interband mixing.

The required ingredients are compatible with established techniques in ultracold-atom optical lattices, including optical Lieb-lattice realization, coherent flat-band preparation, controlled quasimomentum transport, laser-assisted complex tunneling, and real-space density imaging. The geometric response can therefore be inferred from the initial force-induced evolution of the cloud width without reconstructing the full Bloch wave function. In this sense, wave-packet spreading provides a real-space counterpart to anomalous center-of-mass motion: the latter probes Berry curvature, whereas the former can reveal momentum-space variation of the quantum metric. This distinction is particularly useful in flat bands with vanishing Berry curvature, where the metric contribution can be isolated without a competing leading-order transverse drift. More generally, the same framework can be applied to other engineered multiband systems in which band geometry remains nontrivial despite strongly suppressed dispersion.

\appendix

\section{DERIVATION OF THE WAVEPACKET DYNAMICS}
\label{appendix: Appendix A}

Let us consider an $N$-band Bloch Hamiltonian with an isolated flat band, labeled by $n=0$, and remaining dispersive bands. We take the initial wave packet to be constructed entirely from the flat-band eigenstates,
\begin{eqnarray}
    \ket{\Psi(0)}
    &=&
    \int_{\mathrm{BZ}} [d k]\, a_0(\bm{k},0)\ket{\psi_{0\bm{k}}}
    \nonumber \\
    &=&
    \int_{\mathrm{BZ}} [d k]\, a_0(\bm{k},0)
    e^{i\bm{k}\cdot\bm{r}}\ket{u_0(\bm{k})},
    \label{App:intlwf}
\end{eqnarray}
where $\ket{u_0(\bm{k})}$ is the cell-periodic part of the flat-band Bloch eigenstate. At time $t=0$, we turn on a constant external force $\bm F$, corresponding to the perturbation
\begin{eqnarray}
    \delta H = -\bm F\cdot \hat{\bm r}.
    \label{App:Hpert}
\end{eqnarray}

We first derive the action of the position operator in the Bloch eigenstate basis. Taking a momentum derivative of the Bloch state gives
\begin{eqnarray}
    \partial_{k_i}\ket{\psi_{n\bm{k}}}
    &=&
    \partial_{k_i}
    \left(e^{i\bm{k}\cdot\bm r}\ket{u_n(\bm{k})}\right)
    \nonumber \\
    &=&
    i\hat r_i\ket{\psi_{n\bm{k}}}
    +
    e^{i\bm{k}\cdot\bm r}
    \partial_{k_i}\ket{u_n(\bm{k})}.
    \label{App:derivative_1}
\end{eqnarray}
Using the completeness relation
\begin{eqnarray}
    \mathbb{I}(\bm{k})
    =
    \sum_m
    \ket{u_m(\bm{k})}\bra{u_m(\bm{k})},
\end{eqnarray}
we can write
\begin{eqnarray}
    e^{i\bm{k}\cdot\bm r}
    \partial_{k_i}\ket{u_n(\bm{k})}
    =
    -iA^i_n(\bm{k})\ket{\psi_{n\bm{k}}}
    -
    i\sum_{m\neq n}A^i_{mn}(\bm{k})\ket{\psi_{m\bm{k}}},
\end{eqnarray}
where the intraband and interband Berry connections are defined as
\begin{eqnarray}
    A^i_n(\bm{k})
    &=&
    i\braket{u_n(\bm{k})|\partial_{k_i}u_n(\bm{k})},
    \\
    A^i_{mn}(\bm{k})
    &=&
    i\braket{u_m(\bm{k})|\partial_{k_i}u_n(\bm{k})}.
\end{eqnarray}
Substituting this result into Eq.~\eqref{App:derivative_1}, we obtain
\begin{eqnarray}
    \partial_{k_i}\ket{\psi_{n\bm{k}}}
    =
    i\hat r_i\ket{\psi_{n\bm{k}}}
    -
    iA^i_n(\bm{k})\ket{\psi_{n\bm{k}}}
    -
    i\sum_{m\neq n}A^i_{mn}(\bm{k})\ket{\psi_{m\bm{k}}}.
\end{eqnarray}
Therefore, the position operator acts as
\begin{eqnarray}
    \hat r_i\ket{\psi_{n\bm{k}}}
    &=&
    \left(-i\partial_{k_i}+A^i_n(\bm{k})\right)
    \ket{\psi_{n\bm{k}}}
    +
    \sum_{m\neq n}A^i_{mn}(\bm{k})\ket{\psi_{m\bm{k}}}
    \nonumber \\
    &=&
    D^i_n\ket{\psi_{n\bm{k}}}
    +
    \sum_{m\neq n}A^i_{mn}(\bm{k})\ket{\psi_{m\bm{k}}},
    \label{App:posopfinal}
\end{eqnarray}
where
\begin{eqnarray}
    D^i_n=-i\partial_{k_i}+A^i_n(\bm{k})
\end{eqnarray}
is the covariant derivative associated with band $n$.

Let us now evolve the wave function after turning on the perturbation in Eq.~\eqref{App:Hpert}. We work in the interaction picture, where
\begin{eqnarray}
    \ket{\tilde{\Psi}(t)}
    =
    e^{iH_0t/\hbar}\ket{\Psi(t)},
    \label{App:intpicture}
\end{eqnarray}
and the Schr\"odinger equation becomes
\begin{eqnarray}
    i\hbar \partial_t\ket{\tilde{\Psi}(t)}
    =
    \delta H_I(t)\ket{\tilde{\Psi}(t)},
    \label{App:SE}
\end{eqnarray}
with
\begin{eqnarray}
    \delta H_I(t)
    =
    e^{iH_0t/\hbar}\delta H e^{-iH_0t/\hbar}.
\end{eqnarray}
We expand the interaction-picture state in the unperturbed Bloch basis as
\begin{eqnarray}
    \ket{\tilde{\Psi}(t)}
    =
    \sum_n\int_{\mathrm{BZ}}[dk]\,
    a_n(\bm{k},t)\ket{\psi_{n\bm{k}}}.
    \label{App:psitildet}
\end{eqnarray}

\begin{widetext}

Upon substituting Eq.~\eqref{App:psitildet} into Eq.~\eqref{App:SE}, and using Eq.~\eqref{App:posopfinal}, we obtain
\begin{eqnarray}
    \dot{a}_m(\bm{k},t)
    &=&
    \frac{i}{\hbar}F_i
    \left[
    \left(\mathcal{D}^i_m(\bm{k})+t v^i_{m\bm{k}}\right)
    a_m(\bm{k},t)
    +
    \sum_{n\neq m}
    e^{i(E_{m\bm{k}}-E_{n\bm{k}})t/\hbar}
    A^i_{mn}(\bm{k})a_n(\bm{k},t)
    \right],
    \label{App:nonflatevol}
    \\
    \dot{a}_0(\bm{k},t)
    &=&
    \frac{i}{\hbar}F_i
    \left[
    \mathcal{D}^i_0(\bm{k})a_0(\bm{k},t)
    +
    \sum_{n\neq 0}
    e^{i(E_0-E_{n\bm{k}})t/\hbar}
    A^i_{0n}(\bm{k})a_n(\bm{k},t)
    \right],
    \label{App:flatevol}
\end{eqnarray}
where $\bm v_{m\bm{k}}=\hbar^{-1}\nabla_{\bm{k}}E_{m\bm{k}}$ is the group velocity of band $m$, and
\begin{eqnarray}
    \mathcal D^i_m(\bm{k})=i\partial_{k_i}+A^i_m(\bm{k})
    \label{App:newD}
\end{eqnarray}
acts on the momentum-space coefficient $a_m(\bm{k},t)$. We shall use $D^i_n=-i\partial_{k_i}+A^i_n$ when the derivative acts on the Bloch eigenket, and $\mathcal D^i_n=i\partial_{k_i}+A^i_n$ when the corresponding projected operator acts on the momentum-space coefficient after integration by parts. Since the flat band has energy $E_0$ independent of $\bm{k}$, its group velocity vanishes.

We now derive the condition under which leakage from the flat band into the dispersive bands can be neglected. Since the initial wave packet is prepared entirely in the flat band, $a_n(\bm{k},0)=0$ for all $n\neq 0$. To leading order in $|\bm F|$, the amplitude generated in band $n\neq 0$ obeys
\begin{eqnarray}
    \dot{a}_n(\bm{k},t)
    \approx
    \frac{i}{\hbar}F_i
    e^{i(E_{n\bm{k}}-E_0)t/\hbar}
    A^i_{n0}(\bm{k})a_0(\bm{k},0).
    \label{App:amdot}
\end{eqnarray}
Integrating Eq.~\eqref{App:amdot}, we obtain
\begin{eqnarray}
    a_n(\bm{k},t)
    =
    \frac{F_iA^i_{n0}(\bm{k})}{\Delta_{n0}(\bm{k})}
    \left[
    e^{i\Delta_{n0}(\bm{k})t/\hbar}-1
    \right]
    a_0(\bm{k},0),
    \label{App:interband}
\end{eqnarray}
where $\Delta_{n0}(\bm{k})=E_{n\bm{k}}-E_0$ is the energy separation between band $n$ and the flat band. The corresponding interband transition probability is
\begin{eqnarray}
    P_{0\rightarrow n}(\bm{k},t)
    =
    4
    \frac{
    \left|\bm F\cdot \bm A_{n0}(\bm{k})\right|^2
    }
    {
    \Delta_{n0}^2(\bm{k})
    }
    \sin^2
    \left(
    \frac{\Delta_{n0}(\bm{k})t}{2\hbar}
    \right)
    \left|a_0(\bm{k},0)\right|^2 .
    \label{App:transprob}
\end{eqnarray}
Thus, the total leakage probability remains perturbatively small provided
\begin{eqnarray}
     \sum_{n\neq0}
     \int_{\mathrm{BZ}}[dk]\,
     4
     \frac{
     \left|\bm F\cdot \bm A_{n0}(\bm{k})\right|^2
     }
     {
     \Delta_{n0}^2(\bm{k})
     }
     \sin^2
     \left(
     \frac{\Delta_{n0}(\bm{k})t}{2\hbar}
     \right)
     \left|a_0(\bm{k},0)\right|^2
     \ll 1 .
     \label{App:sufficient}
\end{eqnarray}
A simpler, stronger sufficient condition is
\begin{eqnarray}
    \max_{\bm{k}}\sum_{n\neq0}
    \frac{
    \left|\bm F\cdot \bm A_{n0}(\bm{k})\right|
    }
    {
    |\Delta_{n0}(\bm{k})|
    }
    \ll 1.
    \label{App:bandgap}
\end{eqnarray}
A band-summed sufficient criterion may also be expressed in terms of the flat-band quantum metric as
\begin{eqnarray}
    \max_{\bm{k}}
    \frac{
    \sqrt{F_iF_jg^{ij}_0(\bm{k})}
    }
    {
    \Delta_{\min}
    }
    \ll 1,
    \label{App:strong_sufficient}
\end{eqnarray}
where $\Delta_{\min}$ is the minimum energy separation between the flat band and all remaining bands over the relevant region of the Brillouin zone.

Under the negligible interband-leakage condition in Eq.~\eqref{App:strong_sufficient}, the dynamics reduce to the projected flat-band equation
\begin{eqnarray}
    \dot{a}_0(\bm k,t)
    =
    \frac{i}{\hbar}F_i\mathcal{D}^i_0(\bm k)a_0(\bm k,t).
    \label{App:noleakevol}
\end{eqnarray}
This equation has the formal solution
\begin{eqnarray}
    a_0(\bm k,t)
    &=&
    e^{\frac{it}{\hbar}F_i\mathcal{D}^i_0}
    a_0(\bm k,0)
    \label{analytic_soln}
    \\
    &\approx&
    a_0(\bm k,0)
    +
    \frac{it}{\hbar}F_i\mathcal{D}^i_0a_0(\bm k,0)
    +
    \mathcal{O}(t^2),
    \label{App:linear_soln}
\end{eqnarray}
where the second line gives the leading short time expansion. This linear approximation is valid when the first-order correction is small compared with the initial envelope. A sufficient local condition is
\begin{eqnarray}
    \left|
    \frac{t}{\hbar}
    F_i
    \left[
    -\frac{\partial_{k_i}a_0(\bm k,0)}{a_0(\bm k,0)}
    +
    iA^i_0(\bm k)
    \right]
    \right|
    \ll 1 .
    \label{App:lintimecondt}
\end{eqnarray}
For a Gaussian wave packet constructed in the flat band of the Lieb lattice model in Eq.~\eqref{eq:cold_atom_lieb_hamiltonian}, with initial real-space standard deviation $\sigma_r$, this condition reduces to
\begin{eqnarray}
    \frac{\sigma_r|\bm F|t}{\hbar}\ll 1.
    \label{App:gaussian_short_time}
\end{eqnarray}

To characterize the wave-packet dynamics, we study the time evolution of its position moments. In particular, we first derive the expectation values of the first and second position moments for the general time-evolved wave packet in Eq.~\eqref{App:psitildet}, and then specialize to the short time isolated-flat-band limit.

Using Eqs.~\eqref{App:psitildet} and \eqref{App:intpicture}, together with the definition of expectation values, the first position moment is
\begin{eqnarray}
    \langle r_\alpha(t)\rangle
    &=&
    \sum_n\int[dk]\,
    \left[
    t v^\alpha_n(\bm k)\left|a_n(\bm k,t)\right|^2
    +
    a_n^*(\bm k,t)\mathcal{D}^\alpha_n(\bm k)a_n(\bm k,t)
    \right]
    \nonumber \\
    &&+
    \sum_n\sum_{p\neq n}
    \int[dk]\,
    e^{\frac{i}{\hbar}(E_p(\bm k)-E_n(\bm k))t}
    A^\alpha_{pn}(\bm k)
    a_p^*(\bm k,t)a_n(\bm k,t).
    \label{App:mean}
\end{eqnarray}

The corresponding second moment can be written exactly in the compact form
\begin{eqnarray}
    \langle r_\alpha^2(t)\rangle
    =
    \sum_m\int[dk]\,
    \left|
    \left(
    \mathcal D_m^\alpha
    +
    t v_m^\alpha
    \right)
    a_m(\bm k,t)
    +
    \sum_{n\neq m}
    e^{\frac{i}{\hbar}
    (E_m(\bm k)-E_n(\bm k))t}
    A^\alpha_{mn}(\bm k)
    a_n(\bm k,t)
    \right|^2 .
    \label{App:secmoment}
\end{eqnarray}
Here, $\mathcal{D}^i_n=i\partial_{k_i}+A^i_n$ acts on the momentum-space coefficients, in contrast to $D^i_n=-i\partial_{k_i}+A^i_n$, which acts on the Bloch eigenkets before integration by parts. We also use the convention that repeated Roman indices are summed over, while the Greek index $\alpha$ is fixed.

In the isolated flat-band limit, the dispersive-band amplitudes are neglected and the flat-band group velocity vanishes. Equations~\eqref{App:mean} and \eqref{App:secmoment} therefore reduce to
\begin{eqnarray}
    \langle r_\alpha(t)\rangle
    &=&
    \int [dk]\,
    a_0^*(\bm k,t)\mathcal{D}^\alpha_0(\bm k)a_0(\bm k,t),
    \label{App:flat_band_mean}
    \\
    \langle r_\alpha^2(t)\rangle
    &=&
    \int[dk]\,
    \left|\mathcal{D}^\alpha_0(\bm k)a_0(\bm k,t)\right|^2
    +
    \int[dk]\,
    \left|a_0(\bm k,t)\right|^2 g^{\alpha\alpha}_0(\bm k),
    \label{App:flat_band_r2}
\end{eqnarray}
where we have used
\begin{equation}
    g_0^{\alpha\alpha}(\bm k)
    =
    \sum_{m\neq0}
    \left|A^\alpha_{m0}(\bm k)\right|^2
\end{equation}
for the diagonal components of the flat-band quantum metric.

\begin{figure*}[t]
    \centering
    \includegraphics[width=\linewidth]{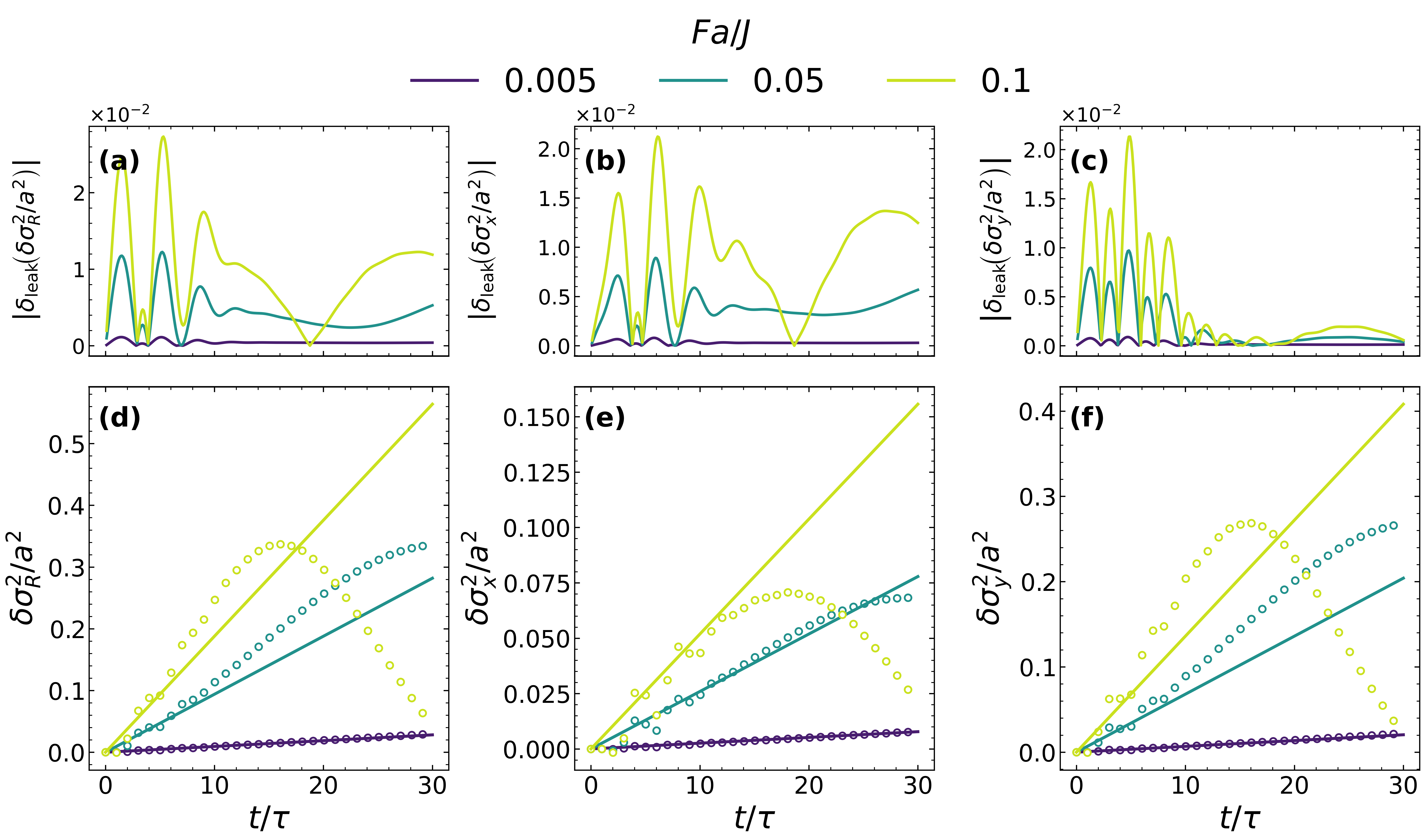}
    \caption{Effect of interband leakage on wave-packet spreading for increasing force strength. (a-c) Leakage-induced correction to the total, \(x\)-resolved, and \(y\)-resolved variances, respectively. (d-f) Corresponding numerical spreading (open circles) compared with the leading order analytical prediction (solid lines). Results are shown for \(Fa/J=0.005,\ 0.05,\) and \(0.1\), with \(m/J=0.25\), \(\lambda_c=1.0\), and \(\sigma_r/a=1.75\).}
    \label{fig:Leakage_force}
\end{figure*}

Using the Berry-curvature convention
\begin{equation}
    \Omega_{\alpha j}
    =
    \partial_{k_\alpha}A^j_0
    -
    \partial_{k_j}A^\alpha_0,
\end{equation}
and substituting the short time solution in Eq.~\eqref{App:linear_soln} into Eqs.~\eqref{App:flat_band_mean} and \eqref{App:flat_band_r2}, we obtain the leading changes
\begin{eqnarray}
    \langle r_\alpha(t)\rangle-\langle r_\alpha(0)\rangle
    &=&
    -\frac{t}{\hbar}
    F_j
    \langle \Omega_{\alpha j}\rangle,
    \label{App:mean_drift}
    \\
    \sigma_{r_\alpha}^2(t)-\sigma_{r_\alpha}^2(0)
    &=&
    \frac{t}{\hbar}
    F_j
    \left[
    -\langle \{\mathcal{D}_0^\alpha,\Omega_{\alpha j}\}\rangle
    +
    2\langle \mathcal{D}_0^\alpha\rangle
    \langle\Omega_{\alpha j}\rangle
    +
    \langle\partial_{k_j}g_0^{\alpha\alpha}\rangle
    \right].
    \label{App:var_drift}
\end{eqnarray}
Here, $\{\, ,\,\}$ denotes the anticommutator and the index $\alpha$ is fixed throughout Eq.~\eqref{App:var_drift}.

\end{widetext}

\begin{figure*}
    \centering
    \includegraphics[width=\linewidth]{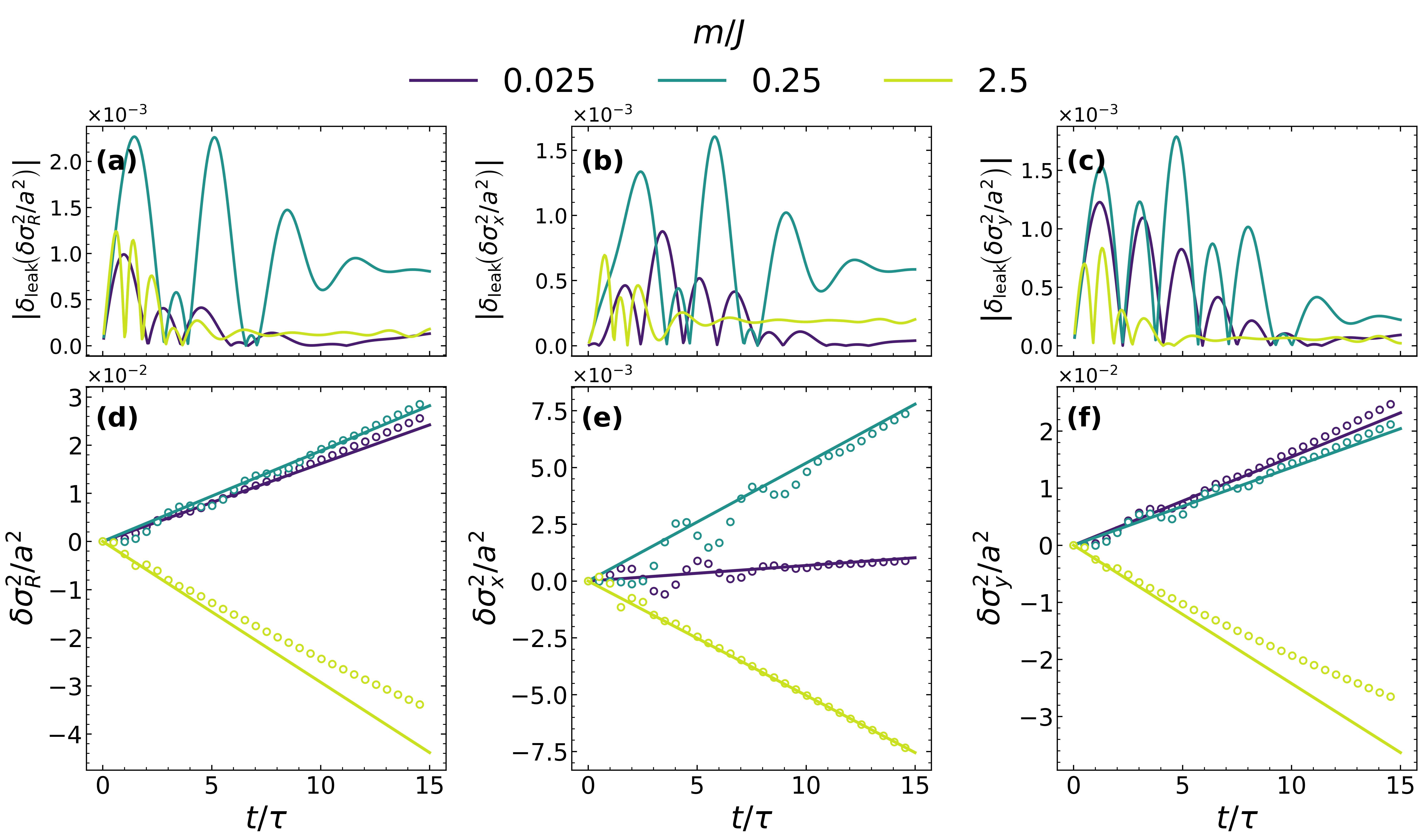}
    \caption{Effect of interband leakage on wave-packet spreading for different complex \(B\)-\(C\) hopping strengths. (a-c) Leakage-induced correction to the total, \(x\)-resolved, and \(y\)-resolved variances, respectively. (d-f) Corresponding numerical spreading (open circles) compared with the leading-order analytical prediction (solid lines). Results are shown for \(m/J=0.025,\ 0.25,\) and \(2.5\), with \(Fa/J=0.01\), \(\lambda_c=1.0\), and \(\sigma_r/a=1.75\).}
    \label{fig:Leakage_gap}
\end{figure*}

\section{Interband leakage and validity of the flat-band approximation}
\label{appendix: Leakage}

Although the initial state is projected onto the flat band of \(H_0\), the applied force does not, in general, preserve this subspace. To quantify departures from the flat-band manifold, we calculate the instantaneous flat-band weight
\begin{equation}
    W_{\mathrm{flat}}(t)
    =
    \langle\Psi(t)|P_0|\Psi(t)\rangle,
    \label{eq:time_flat_weight}
\end{equation}
where \(P_0\) is the projector onto the flat band of the force-free Hamiltonian defined in Eq.~\eqref{eq:flatband_projector}. The corresponding norm leakage into the complementary dispersive-band subspace is
\begin{equation}
    \mathcal{L}(t)
    =
    1-W_{\mathrm{flat}}(t).
    \label{eq:leakage}
\end{equation}

To quantify the influence of the contributions from the dispersive bands on the real-space spreading, we also compare the variance obtained from the full state with that obtained from the normalized flat-band-projected state in Eq.~\eqref{eq:normalized_time_projected_state}. We define the leakage-induced correction to each variance component as
\begin{equation}
    \delta_{\mathrm{leak}}
    \left(
    \frac{\delta\sigma_i^2}{a^2}
    \right)
    =
    \frac{
    \delta\sigma_{i,\mathrm{full}}^2
    -
    \delta\sigma_{i,\mathrm{flat}}^2
    }{a^2},
    \qquad
    i=x,y,R,
    \label{eq:leakage_variance_correction}
\end{equation}
where \(R\) denotes the total variance, \(\sigma_R^2=\sigma_x^2+\sigma_y^2\). This observable correction is distinct from the norm leakage \(\mathcal L(t)\). In particular, position moments can contain interference between the flat-band and dispersive-band components, so the correction to the variance need not scale directly with the leaked probability.

Figure~\ref{fig:Leakage_force} shows the effect of increasing the applied force while keeping the remaining parameters fixed. Panels (a)-(c) show the magnitude of the leakage-induced correction in Eq.~\eqref{eq:leakage_variance_correction} for the total, \(x\)-resolved, and \(y\)-resolved variances, respectively, while panels (d)-(f) compare the corresponding numerical spreading with the leading-order analytical prediction. The oscillatory structure of the leakage-induced correction is reflected in the deviations of the numerical spreading from the linear prediction, particularly at larger force. For weak force, however, the correction remains small and the spreading follows the leading-order behavior over the short time regime
\[
\frac{t}{\tau}\ll\frac{1}{\sigma(Fa/J)}.
\]
The resolved components further show that a force applied along \(y\) can modify both \(\sigma_x^2\) and \(\sigma_y^2\).

Figure~\ref{fig:Leakage_gap} shows the corresponding comparison when the complex \(B\)-\(C\) hopping \(m\) is varied at fixed applied force. Changing \(m\) modifies both the separation of the flat band from the dispersive bands and the associated interband matrix elements, leading to different leakage-induced corrections. Despite these changes, the numerical spreading remains close to the leading-order prediction over the short time interval where the correction is small. The \(x\)- and \(y\)-resolved panels also show that for the force applied along \(y\), it selects the \(k_y\) derivative of the metric, but does not restrict the real-space spreading to the \(y\) direction alone.

Taken together, Figs.~\ref{fig:Leakage_force} and \ref{fig:Leakage_gap} show that the oscillatory corrections associated with interband mixing remain superimposed on the leading geometric spreading. Interband leakage therefore provides a diagnostic for identifying the regime over which the projected flat-band approximation and the leading-order analytical prediction remain accurate.

\section{Real space origin of the complex \texorpdfstring{$B$-$C$}{B-C} hopping}
\label{app:raman_implementation}

The nearest-neighbor tunneling terms in Eqs.~\eqref{eq:real_space_AB} and \eqref{eq:real_space_AC} correspond to the standard optical Lieb lattice. The nontrivial ingredient required to open a gap while preserving the exactly flat middle band is the complex next-nearest-neighbor tunneling between the \(B\) and \(C\) sublattices.

The desired real-space Hamiltonian is
\begin{align}
    H_{BC}
    &=
    im\sum_{\mathbf R}
    \Big(
        b_{\mathbf R}^{\dagger}c_{\mathbf R}
        -
        b_{\mathbf R}^{\dagger}c_{\mathbf R-\hat{y}} \nonumber \\
        &-
        b_{\mathbf R}^{\dagger}c_{\mathbf R+\hat{x}}
        +
        b_{\mathbf R}^{\dagger}
        c_{\mathbf R+\hat{x}-\hat{y}}
    \Big)
    +\mathrm{H.c.},
    \label{App:HBC}
\end{align}
which assigns the four diagonal hoppings the amplitudes
\[
\{+im,-im,-im,+im\}.
\]
The alternating sign structure is essential for generating the momentum dependence of the Bloch Hamiltonian.

Using the Fourier transformation
\begin{equation}
    \alpha_{\mathbf R}
    =
    \frac{1}{\sqrt{N_{\mathrm{uc}}}}
    \sum_{\mathbf k}
    e^{i\mathbf k\cdot
    (\mathbf R+\boldsymbol{\tau}_{\alpha})}
    \alpha_{\mathbf k},
    \qquad
    \alpha=b,c,
\end{equation}
with
\(\boldsymbol{\tau}_b=(a/2,0)\) and
\(\boldsymbol{\tau}_c=(0,a/2)\),
the four hopping processes contribute the phase factors
\begin{eqnarray}
\sum_{\bm{R}} b_{\bm{R}}^\dagger c_{\bm{R}}
&=&
\sum_{k_x,k_y}
e^{-ik_xa/2+ik_ya/2}
b_{\bm{k}}^\dagger c_{\bm{k}}, \label{app:HBCT1}
\\
\sum_{\bm{R}} b_{\bm{R}}^\dagger
c_{\bm{R}-a\hat{\bm{y}}}
&=&
\sum_{k_x,k_y}
e^{-ik_xa/2-ik_ya/2}
b_{\bm{k}}^\dagger c_{\bm{k}}, \label{app:HBCT2}
\\
\sum_{\bm{R}} b_{\bm{R}}^\dagger
c_{\bm{R}+a\hat{\bm{x}}}
&=&
\sum_{k_x,k_y}
e^{ik_xa/2+ik_ya/2}
b_{\bm{k}}^\dagger c_{\bm{k}}, \label{app:HBCT3}
\\
\sum_{\bm{R}} b_{\bm{R}}^\dagger
c_{\bm{R}+a\hat{\bm{x}}-a\hat{\bm{y}}}
&=&
\sum_{k_x,k_y}
e^{ik_xa/2-ik_ya/2}
b_{\bm{k}}^\dagger c_{\bm{k}}. \label{app:HBCT4}
\end{eqnarray}
respectively. Thus substituting Eqs.~\eqref{app:HBCT1}-~\eqref{app:HBCT4} into the BC hopping term in Eq.~\eqref{App:HBC} and simplifying gives
\begin{eqnarray}
    H_{BC}
    =
    4im
    \sum_{k_x,k_y}\sin\left(\frac{k_xa}{2}\right)
    \sin\left(\frac{k_ya}{2}\right)b_{\bm k}^\dagger c_{\bm k} + \mathrm{H.c}.
\end{eqnarray}
The resulting momentum-space hopping is therefore
\begin{equation}
    H_{BC}(\mathbf k)
    =
    4im
    \sin\left(\frac{k_xa}{2}\right)
    \sin\left(\frac{k_ya}{2}\right),
\end{equation}
which is the momentum-space term appearing in Eq.~\eqref{eq:cold_atom_lieb_hamiltonian}.

Experimentally, the complex tunneling required by the model could be implemented by adapting established Raman- and laser-assisted tunneling schemes in tilted optical lattices. Such techniques have been demonstrated to restore otherwise suppressed tunneling while imprinting controllable Peierls phases on the effective hopping amplitudes~\cite{PhysRevLett.107.255301,PhysRevLett.111.185301,PhysRevLett.111.185302}. Realizing the present model would require engineering the relative phases of the four diagonal \(B\)-\(C\) tunneling processes to produce the pattern
\[
\{+i,-i,-i,+i\}.
\]
The specific four-bond pattern is therefore the implementation required by the present Lieb-lattice model, while the underlying phase-controlled tunneling mechanism is established experimentally.

\bibliography{References}
\end{document}